\documentclass[preprint,aps,prd,showpacs,preprintnumbers]{revtex4-1}

\usepackage{graphicx}
\usepackage{amsmath,amsfonts,amssymb}
\usepackage{latexsym}
\usepackage[utf8]{inputenc}

\begin{document}
\title{Does the diffusion DM-DE interaction model solve cosmological puzzles? 
}

\author{Marek Szyd{\l}owski}
\email{marek.szydlowski@uj.edu.pl}
\affiliation{Astronomical Observatory, Jagiellonian University, Orla 171, 30-244 Krakow, Poland}
\affiliation{Mark Kac Complex Systems Research Centre, Jagiellonian University, {\L}ojasiewicza 11, 30-348 Krak{\'o}w, Poland}
\author{Aleksander Stachowski}
\email{aleksander.stachowski@uj.edu.pl}
\affiliation{Astronomical Observatory, Jagiellonian University, Orla 171, 30-244 Krakow, Poland}

\begin{abstract}
We study dynamics of cosmological models with diffusion effects modeling dark matter and dark energy interactions. We show the simple model with diffusion between the cosmological constant sector and dark matter, where the canonical scaling law of dark matter $(\rho_{dm,0}a^{-3}(t))$ is modified by an additive $\epsilon(t)=\gamma t a^{-3}(t)$ to the form $\rho_{dm}=\rho_{dm,0}a^{-3}(t)+\epsilon(t)$. We reduced this model to the autonomous dynamical system and investigate it using dynamical system methods. This system possesses a two-dimensional invariant submanifold on which the DM-DE interaction can be analyzed on the phase plane. The state variables are density parameter for matter (dark and visible) and parameter $\delta$ characterizing the rate of growth of energy transfer between the dark sectors. A corresponding dynamical system belongs to a general class of jungle type of cosmologies represented by coupled cosmological models in a Lotka-Volterra framework. We demonstrate that the de Sitter solution is a global attractor for all trajectories in the phase space and there are two repellers: the Einstein-de Sitter universe and the de Sitter universe state dominating by the diffusion effects. We distinguish in the phase space trajectories, which become in good agreement with the data. They should intersect a rectangle with sides of $\Omega_{m,0}\in [0.2724, 0.3624]$, $\delta \in [0.0000, 0.0364]$ at the 95\% CL. Our model could solve some of the puzzles of the $\Lambda$CDM model, such as the coincidence and fine-tuning problems. In the context of the coincidence problem, our model can explain the present ratio of $\rho_{m}$ to $\rho_{de}$, which is equal $0.4576^{+0.1109}_{-0.0831}$ at a 2$\sigma$ confidence level.
\end{abstract}

\maketitle

\section{Introduction}

Thanks to astronomical observations the modern cosmology has elaborated the concept of the standard cosmological model called the cold dark matter model with the cosmological constant ($\Lambda$CDM model). From a methodological point of view cosmology has achieved the similar status as the particle physics with its standard model of particles. The $\Lambda$CDM model acts as an effective theory describing the Universe from redshift $z = 0$ (today) to $z = 10^9$ (the epoch of primordial nuclesynthesis). In the very structure of the $\Lambda$CDM model there are essentially two components: matter (dark matter and baryonic matter) and dark energy. It is assumed that the universe is spatially homogeneous and isotropic, and that its evolution is governed by Einstein field equations with the energy momentum tensor for the ideal fluid satisfying the barotropic equation of state $p = p(\rho)$, where $\rho$ is the energy density of fluid. From the observational point of view it is convenient to use dimensionless density parameters $\Omega_i$, defined as the fractions of critical density $3 H_0^2$ which corresponds to a flat model. These parameters are observables which can be determined from astronomical observations \cite{Ade:2015xua}. In the $\Lambda$CDM model it is assumed that all fluids are non-interacting. This implies that the densities of baryonic matter and dark matter are scaling with respect to redshift as $(1 + z)^3$ and dark energy has constant density.

The natural interpretation of the cosmological constant is to treat it as the energy of the quantum vacuum \cite{Weinberg:1988cp}. The cosmological model with the cosmological constant term and pressureless matter fits well to the observation data of measurements of SNIa luminosity distance as a function of redshift \cite{Perlmutter:1997zf} and observations microwave relic radiation (WMAP, Planck). Measurements of large-scale structures also remain consistent with the $\Lambda$CDM model. Although the $\Lambda$CDM model describes well the present Universe the nature of its basic constituents (dark energy and dark energy) remains unknown. So dark energy and dark matter are like some useful fictions in the terminology of Nancy Cartwright \cite{Cartwright:1983hl}. Comparing the value of the cosmological constant required to explain the effect of the accelerated expansion of the Universe observations of distant supernovae type SNIa with the value of the cosmological constant interpreted as the energy of the quantum vacuum,  we get the most incredible gap in history of physics $\rho_{vac} / \rho_{\Lambda} = 10^{143}$. In this context, the question is born why the value of the cosmological constant is the small, why not simply a zero? This is the known problem of the cosmological constant. In the model under the consideration, the comparison of $\rho_{vac} / \rho_{\Lambda}$ still gives 83 order of magnitude from the measurement value because $\rho_{vac}/\rho_{\Lambda}\simeq10^{60}$.

Another problem related closely with the problem of the cosmological constant is the coincidence problem \cite{Peebles:1987ek}. This problem is caused by the lack of explanation of why in today's era of the density of dark matter and dark energy are comparable although it is assumed that they have different periods of recombination. In this paper we construct a cosmological model in which it is assumed that the process of interaction between sectors of dark matter and dark energy is continuous. Relativistic diffusion describes the transfer of energy to the sector of dark matter. As a result, we go beyond the standard model assuming from the outset that dark matter and dark energy interact. This effect is described by the running cosmological constant and the modification of the standard scaling law of the dark matter density.

If we assume that general relativity is an effective theory which can be extrapolated to the Planck epoch, then the interpretation of the cosmological constant parameter appeared in the $\Lambda$CDM model as a vacuum energy seems to be natural. By equating this density to the energy density of the zero point energy that is left in a volume after removing all particles, then we obtain that its value is about 120 orders of magnitude higher than the corresponding value required for explanation of acceleration of the Universe in current epoch. In the Universe with such a high value of cosmological constant (dark energy) we have a rapid inflation and galaxies would have no time to form. The lack of explanation of this difference is called the cosmological constant problem.

Its solution can be possible if we can find some physical mechanism lowering dramatically this value during the cosmic evolution. Of course this process should be defined in a covariant way following general relativity principles.

Our hypothesis is that diffusion cosmology can offer the possibility of obtaining a low value of the cosmological constant today because the effects of diffusion effectively produce the running cosmological constant.

We study how a value of effective running cosmological constant parameter changes during the cosmic evolution and for late time is going to be a small constant value.

From the astronomical observations of distant supernovae SNIa and measurements of CMB by Planck, measurement of BAO and other astronomical observations we obtain that the present value of the energy densities of both dark energy and dark matter are of the same order of magnitude \cite{Velten:2014nra}. If we assume that standard cosmological model ($\Lambda$CDM model) is an adequate way to describe the cosmic evolution, then the value $\rho_{de}/\rho_{dm}$ will depend on the cosmological time or redshift and the question arises: Why are two quantities with different time recombination comparable at the present epoch? It is called the cosmic coincidence problem.

We are looking for some physical relativistic mechanism which gives rise to this coincidence observed for the current Universe. In the opposite case very special initial conditions are required for its realization (fine tunning problem). In the framework of diffusion cosmology our investigation of this problem shows that while the values of dark matter and dark energy densities are comparable today they were significantly different in the past history of the Universe. Because the diffusion effects act effectively as fluids which interact each other during the cosmic evolution. In the consequence dark energy is running and a canonical rule of scaling dark matter $\rho_m$ proportional to $a^{-3}$ is adjusted.

The main aim of our paper is to demonstrate how the coincidence problem can be naturally solved in the framework of diffusion cosmology. The interacting dark energy models have been considered by many authors in the context of this problem. One of reasons to study these models is to solve the cosmic coincidence problem \cite{Cai:2004dk,Berger:2007iw,Copeland:2006wr,Pavon:2005yx,Steinhardt:1997ccp}. To this aim different ad-hoc proposed models of an interacting term were postulated {\'a} prori. In these models the covariance of general relativity is usually violated and therefore they have limited application to cosmology. In the present work we consider a unique relativistic diffusion model where an interaction mechanism is motivated physically.

In the study of evolutional scenarios of the model under consideration we apply  the dynamical systems methods \cite{Perko:2001de}. Our model belongs to a general class of jungle type of cosmologies represented by coupled cosmological models in a Lotka-Volterra framework \cite{Perez:2013zya}. 

The crucial role in the organization of the phase space plays the critical point located inside the physical region. The possible bifurcation of this point are studied in details for extracting variability of DE and DM density as the function of the cosmological time. It is interesting that at this critical point $\rho_{dm} \propto \rho_{de}$ (scaling type solution).

\section{Friedmann equation for diffusion interacting of dark matter with dark energy}

Haba et al. postulated a particular model of an energy-momentum exchange between DM and DE sectors, while a baryonic sector was preserved \cite{Haba:2016swv}. In this paper we reconsider this model in the light of aforementioned cosmological problems.

We assume Einstein equations in the form
\begin{equation}
R^{\mu \nu}-\frac{1}{2}g^{\mu\nu} R = T^{\mu\nu},
\end{equation}
where $g^{\mu\nu}$ is the metric, $R^{\mu\nu}$ is the Ricci tensor. In this paper we use the natural units $8\pi G=c=1$.

Because of a cosmological application we assume that the universe has topology $R\times M^3$, where $M^3$ is homogeneous and isotropic space. Then the spacetime metric depends only on one function of the cosmic time $t$--the scale factor $a(t)$. Additionally for simplicity we also assume flatness ($k=0$) of sections $t=\text{const}$. We decompose the energy-momentum tensor on two parts
\begin{equation}
T^{\mu \nu} = T_{de}^{\mu \nu} + T_{dm}^{\mu\nu}.
\end{equation}

We assume the conservation of the total energy momentum, which gives
\begin{equation}
-\nabla_\mu T_{de}^{\mu \nu} = \nabla_\mu T_{dm}^{\mu \nu}\equiv 3\kappa^2 J^{\nu},\label{con}
\end{equation}
where $\kappa^2$ is the diffusion constant and $J^{\nu}$ is the current, which represents a flow of stream of particles.

We also assume that energy density of the dark matter consisting of particles of mass $m$ is transfered by a diffusion mechanism in an environment described by a perfect fluid. There is only unique diffusion which is relativistic invariant and preserves the particle-mass $m$ \cite{Dudley:1965lm}. The corresponding energy-momentum satisfies the conservation law (\ref{con}). 

The Friedman equation in the FRW metric with baryonic matter, dark matter and dark energy reads
\begin{equation}
3H^2  =\rho_b+\rho_{dm}+ \rho_{de},
\end{equation}
where  $\rho_{dm}$ and $\rho_{\text{de}}$  are determined by relations
\begin{gather}
\rho_{\text{dm}} = \rho_{b,0}a^{-3}+\rho_{dm,0}a^{-3}+\gamma (t-t_0)a^{-3}, \\
\rho_{\text{de}} = \rho_{de}(0)-\gamma\int^{t}a^{-3}dt.
\end{gather}

The current $J^\mu$ in equation (\ref{con}) is conserved \cite{Haba:2010qj,Calogero:2011re,Calogero:2012kd}
\begin{equation}
\nabla_\mu J^\mu = 0.
\end{equation}
The above conservation condition for the FRW metric reduces to
\begin{equation}
J^0=\gamma/3\kappa^2 a^{-3}
\end{equation}
with a positive constant $\gamma$ which can be computed from the phase space distribution $\Omega(p, x)$ of diffusing particles \cite{Haba:2016swv}.

The condition (\ref{con}) after calculation of divergence reduces to the continuity conditions for energy density of both matter and dark energy
\begin{align}
\dot\rho_{m}&=-3H\rho_{m}+\gamma a^{-3},\label{darkmatter}\\
\dot\rho_{de}&=-\gamma a^{-3},\label{darkenergy}
\end{align}
where we assume the equation of state for dark energy as $p_{de}=-\rho_{de}$ and for matter as $p_{m}=0$; a dot denotes differentiation with respect to the cosmological time $t$. Equation (\ref{darkmatter}) can be rewritten as
\begin{equation}
a^{-3}\frac{d}{dt}(\rho_{m} a^3)=\gamma a^{-3}\Leftrightarrow \frac{d}{dt}(E)=\gamma,\label{energy}
\end{equation}
where $E$ is the total energy of matter in the comoving volume $V\sim a^3$. From relation (\ref{energy}), we can obtain that $E=\gamma(t-t_0)$.

In our paper \cite{Haba:2016swv} we considered one unique model of an energy transfer from dark energy (DE) to dark matter (DM) with the diffusive interaction in the dark sector where DE and DM can be treated as ideal fluids. Particles are scattering in an environment of other particles. If we assume  that the subsequent scattering events are independent, the particle motion is described by a Markov process. In order, the assumption that the energy of the particle remains finite leads to the conclusion that the Markov process must be a diffusion.

Therefore diffusion is in some sense unique because there is only one diffusion which is relativistic invariant and preserves the mass \cite{Dudley:1965lm,Franchi:2007rd,Haba:2008uy}. In consequence the interaction between DM and DE fluids is defined in a unique way.

\section{Diffusion cosmology}

In the investigation, the dark energy and dark matter interaction plays role in continuity equation. This equation are special case of jungle cosmological models \cite{Perez:2013zya}.
We assume that $\rho_m=\rho_b+\rho_{dm}$, where $\rho_{b}$ is density of baryonic matter and $\rho_{dm}$ is density of dark matter. The equation of state for dark energy is expressed by $p_{de}=-\rho_{de}$ in our model, where $p_{de}$ is pressure of dark energy and the equation of state for matter is given by $p_{m}=0$, where $p_m$ is pressure of matter.

The Friedmann equation is expressed in the following form
\begin{equation}
3H^2=\rho_{b,0}a^{-3}+\rho_{dm,0}a^{-3}+\gamma (t-t_0)a^{-3}+\rho_{de}(0)-\gamma\int^{t}a^{-3}dt,\label{friedmann}
\end{equation}
where $\rho_{b,0}a^{-3}\equiv\rho_b$, $\rho_{dm,0}a^{-3}+\gamma (t-t_0)a^{-3}\equiv\rho_{dm}$, $\rho_{de}(0)-\gamma\int^{t}a^{-3}dt\equiv\rho_{de}$.
From the Friedmann formula we get a condition
\begin{equation}
1=\Omega_{m}+\Omega_{de},\label{condition}
\end{equation}
where $\Omega_{m}=\frac{\rho_{m}}{3H^2}$ and $\Omega_{de}=\frac{\rho_{de}}{3H^2}$ are dimensionless density parameters.

We can obtain equations (\ref{darkmatter})-(\ref{friedmann}) in the form of the dynamical system $x'=f_{x}(x,y,\delta)$,  $y'=f_{y}(x,y,\delta)$ and $\delta'=f_{\delta}(x,y,\delta)$, where $x=\Omega_{m}$, $y=\Omega_{de}$, $\delta=\frac{\gamma a^{-3}}{H\rho_{m}}$ and $'\equiv\frac{d}{d\ln{a}}$ is a differentiation with respect to the reparametrized time $\ln a(t)$. For these variables, the dynamical system is in the following form
\begin{align}
x'&=x(-3+\delta+3x),\label{dyn1}\\
y'&=x(-\delta+3y),\\
\delta'&=\delta(-\delta+\frac{3}{2}x).\label{dyn3}
\end{align}
From equation (\ref{condition}), we have the following relation 
\begin{equation}
x+y=1.
\end{equation}
Then dynamical system (\ref{dyn1})-(\ref{dyn3}) is reduced to the two-dimension dynamical system.

\begin{figure}
	\centering
	\includegraphics[width=0.7\linewidth]{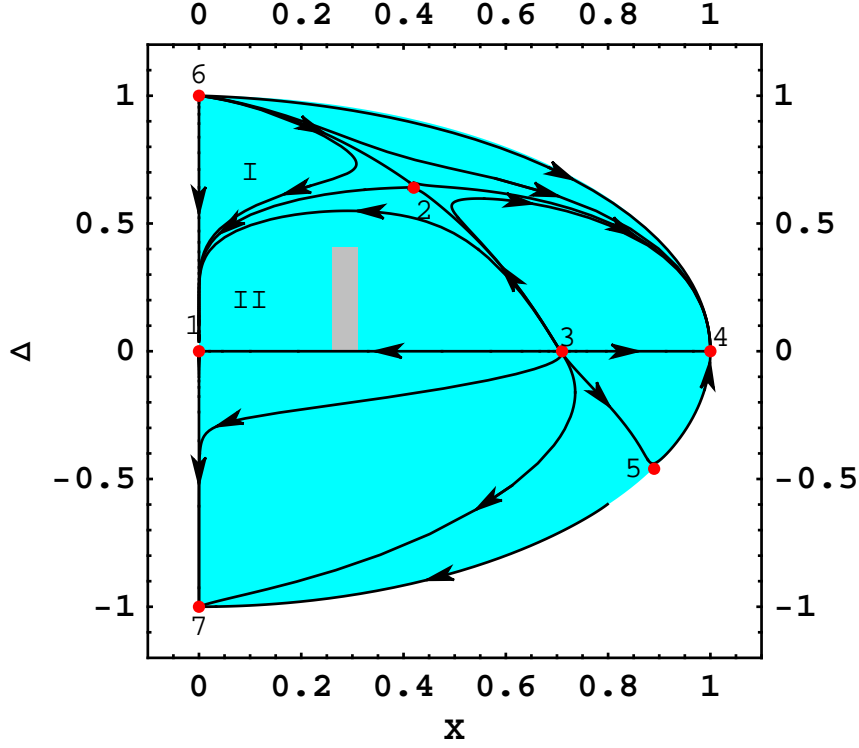}
	\caption{Phase portrait of dynamical system
		$x'=x(\delta+3x-3)$, $\delta'=\delta(-\delta+\frac{3}{2}x)$,
		where $x=\Omega_{m}=\frac{\rho_{m}}{3 H^2}$, $\delta=\frac{\gamma a^{-3}}{H \rho_{m}}$ and $'\equiv\frac{d}{d\ln a}$ on the Poincar{\'e} sphere coordinates are $X=\frac{x}{\sqrt{x^2+y^2+z^2}}$, $\Delta=\frac{\delta}{\sqrt{x^2+y^2+z^2}}$.
		Critical point (1) represents the de Sitter universe--a global attractor for all physical trajectories. Critical point (2) represents the scaling universe. Critical point (3) represents the Einstein-de Sitter universe. Critical points (4) and (5) represent the static universe. Critical point (6) represents the de Sitter universe. The gray region represents the domain of the present value of $X$ and $\Delta$, which is distinguished by astronomical data. Let us note trajectories lie in the domain with $\Delta<0$ represent the contracting model but there is no symmetry with respect to the $\Delta$-axis. At critical point (6), energy density of baryonic matter is negligible as well as density of dark matter and only effects of the relativistic diffusion are important.}  
	\label{fig:fig11}
\end{figure}

For analysis of the critical points in the infinity, we use the Poincar{\'e} sphere. Let us introduce new variables in which one can study dynamical behavior at infinity
\begin{equation}
X=\frac{x}{\sqrt{x^2+\delta^2}}, \quad \Delta=\frac{\delta}{\sqrt{x^2+\delta^2}}.\label{eq:var-xd}
\end{equation}
For these variables we get the dynamical system
\begin{align}
X'&=X\left[-\Delta^2(\frac{3}{2}X-\Delta)+(1-X^2) (3X+\Delta-3\sqrt{1-X^2-\Delta^2})\right],\label{poincare1} \\
\Delta'&=\Delta\left[(1-\Delta^2)(\frac{3}{2}X-\Delta)-X^2 (3X+\Delta-3\sqrt{1-X^2-\Delta^2})\right],\label{poincare2}
\end{align}
where $'\equiv \sqrt{1-X^2-\Delta^2}\frac{d}{d\tau}$. Critical points, for equation (\ref{poincare1})-(\ref{poincare2}), are presented in Table~\ref{table:1}. The phase portrait for the dynamical system (\ref{poincare1})-(\ref{poincare2}) is presented in Figure~\ref{fig:fig11}.

In the phase portrait there is an interesting class of trajectories labeled as ``I'', starting from the critical point 6 and approaching the de Sitter state. Because the diffusion has the physical sense only for interval $t>t_0$ the corresponding cosmological solution should be cut off for any $t<t_0$ from the de-Sitter solution. Hence, we obtain that $a(t_0)$ is positive number, i.e. solution which represent critical point (6) is non singular. All trajectories starting from the de-Sitter state can be treated as a models of extended idea of emergent cosmology \cite{Guendelman:2014bva} \cite{Bag:2014tta}.

While in the standard emergent cosmology universe is starting from the static Einstein model, trajectories of type I are simple realization of extended idea of emergent Universe in which Universe is starting rather from the stationary state.

The results of our previous paper \cite{Haba:2016swv} show that the density parameter for total (dark and visible) $x$ and dimensionless parameter $\delta$ are constrained to $x\in(0.2724,\text{ }0.3624)$, $\delta \in (0.0000,\text{ }0.0364)$ at the $95\%$ confidence level. This domain is represented in the phase space by a shaded rectangle. Only these trajectories which intersect this domain are in good agreement with the observation at a 2$\sigma$ confidence level. Therefore observation favored the cosmological models starting from the Einstein-de Sitter solution and going toward the de Sitter attractor (trajectory II on the phase portrait).

Note that on the phase portrait there are trajectories labeled as ``I'' starting from the de Sitter state and approaching the de Sitter state at late times. They are going toward a saddle point--representing a non-singular solution. However all these trajectories do not intersect the rectangle and therefore they are not favored by observation.

The saddle point in the phase space is representing the Milne universe (see Table~\ref{table:1}). Therefore, the interacting term is proportional to $t^{-3}$ and consequently is of the form $\rho_{\text{dm}} = \Lambda_{\text{bare}} + \alpha^2 t^{-2}$. The cosmological model with such a parametrization of dark energy was studied by Szydlowski and Stachowski \cite{Szydlowski:2015bwa,Szydlowski:2015rga}.

\begin{figure}
	\centering
	\includegraphics[width=0.7\linewidth]{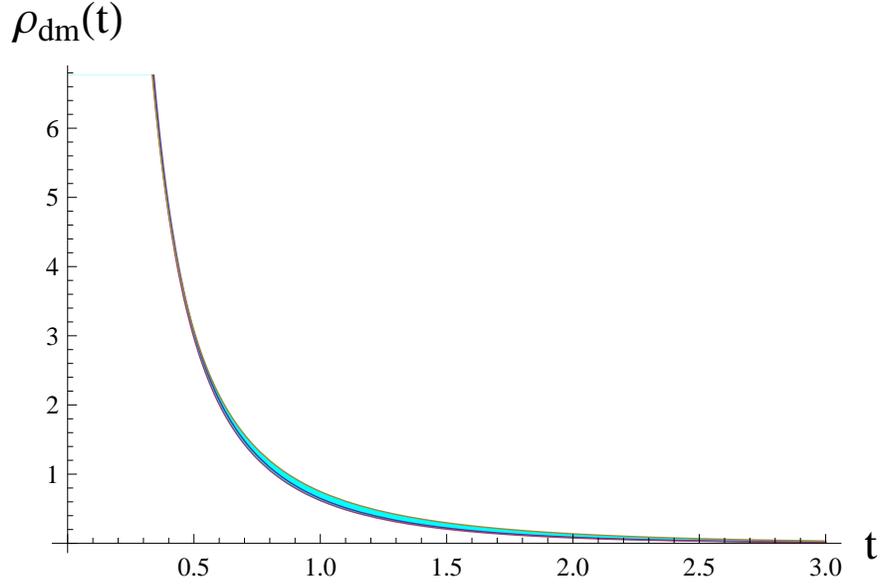}
	\caption{The evolution of dark matter energy density for trajectories of type II (for the best fitted values of model parameter together with confidence level at $95\%$). Dark matter $\rho_{dm}$ is expressed in [100$\times$km/(s Mpc)]$^2$. We choose (s Mpc)/(100$\times$km) as a unit of the cosmological time $t$.
		}
	\label{fig:fig2a}
\end{figure}

\begin{figure}
	\centering
	\includegraphics[width=0.7\linewidth]{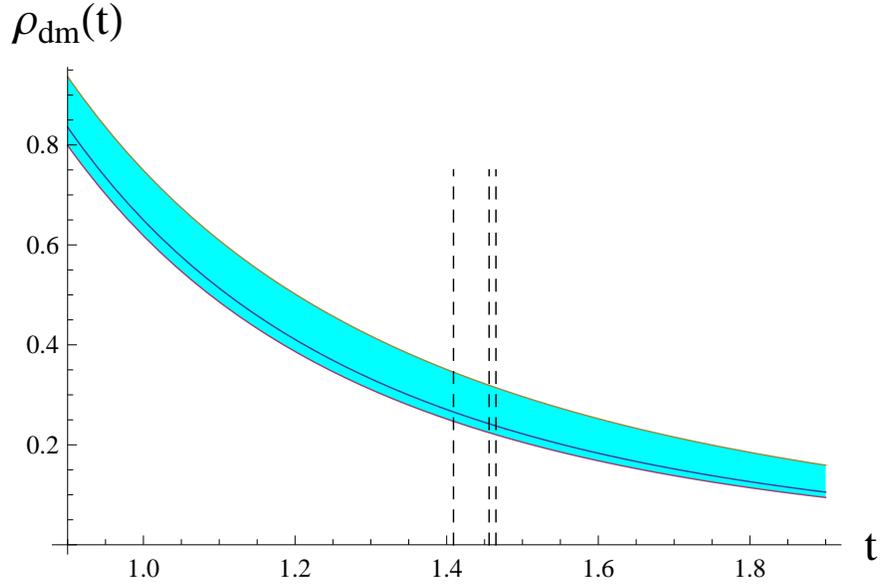}
	\caption{The evolution of dark matter energy density for trajectories of type II (for the best fitted values of model parameter together with confidence level at $95\%$ for the present epoch). Dark matter $\rho_{dm}$ is expressed in [100$\times$km/(s Mpc)]$^2$. We choose (s Mpc)/(100$\times$km) as a unit of the cosmological time $t$. The value of the age of the Universe for the best fit with errors are presented by the dashed lines.}
	\label{fig:fig2b}
\end{figure}

Figures \ref{fig:fig2a} and \ref{fig:fig2b} present the evolution of dark matter $\rho_{dm}$ as a function of the cosmological time $t$ for trajectories of type II. The evolution of the cosmological time for matter is determined by the following formula
\begin{equation}
	\rho_{m}(t)=\rho_{m,0}a(t-t_0)^{-3}+\gamma (t-t_0) a(t-t_0)^{-3}.\label{matter}
\end{equation}

The addictive form of the scaling relation for dark matter (\ref{matter}) suggests that dark matter consists of two components: the first term scaling like $\rho_{m,0} a(t)^{-3}$ and the seconf term scaling like $\gamma t a(t)^{-3}$. The latter describes an amount of dark energy density which is transfered to dark energy sector by the diffusion process. In the unit of $\Omega_{\text{total}}$ the canonically scaling dark matter is $25.23\%$ while transfered dark energy is about $1.35\%$. The amount of transfered dark energy today is of the order $\gamma T$ where $T$ is the age of the Universe.

In consequence of equation (\ref{matter}), $\delta(t)$ can be rewritten as
\begin{equation}
\delta(t)=\frac{1}{H\left(\frac{\rho_{m,0}}{\gamma}+t-t_0\right)}.
\end{equation}
Therefore at the present epoch we have
\begin{equation}
\delta(T)=\frac{1}{H_0 \left(\frac{\rho_{m,0}}{\gamma}+T\right)}, 
\end{equation}
where $t_0=0$ and $t=T$ is the present age of the Universe. Note that while at late time $\delta(t)=\sqrt{\frac{3}{\Lambda}}\frac{1}{t}$ for small time $\delta(t)=\frac{3\gamma}{2\rho_{m,0}}(t-t_0)$.

If we give $\gamma=0$ in equation (\ref{matter}) then $\rho_m$ is scaling in the canonical way. From equation (\ref{matter}) one can simply obtain that the density of dark matter is
\begin{equation}
\rho_{dm}=(\rho_{m,0}-\rho_{b,0})a(t-t_0)^{-3}+\gamma (t-t_0) a(t-t_0)^{-3}.
\end{equation}

\begin{figure}
	\centering
	\includegraphics[width=0.7\linewidth]{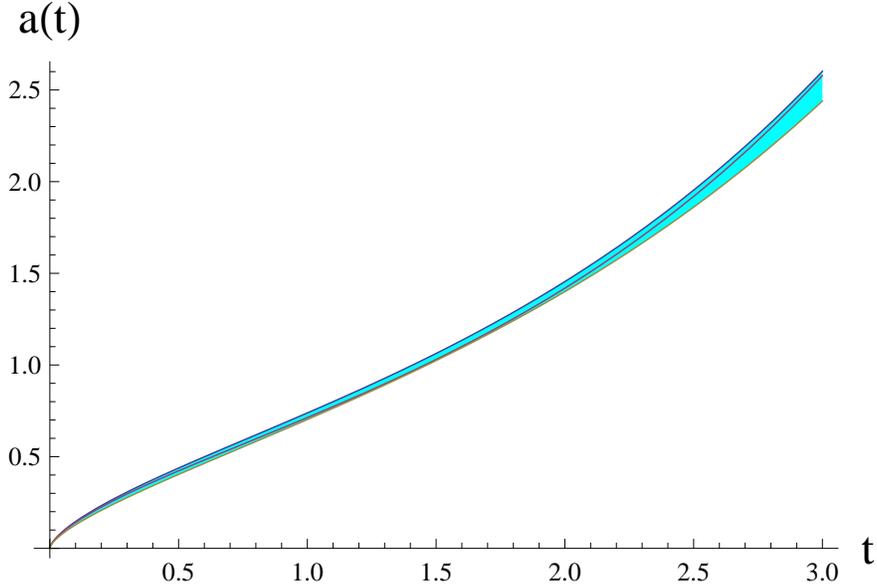}
	\caption{Diagram of scale factor as $a$ function of cosmological time $t$ for trajectories of type II (for the best fitted values of model parameter together with confidence level at $95\%$). For the present epoch $T$ $a(T)=1$. A universe is starting from the initial singularity toward a de Sitter universe. This type of behavior is favored by the observational data. We choose (s Mpc)/(100$\times$km) as a unit of the cosmological time $t$.}
	\label{fig:fig3}
\end{figure}

\begin{figure}
	\centering
	\includegraphics[width=0.7\linewidth]{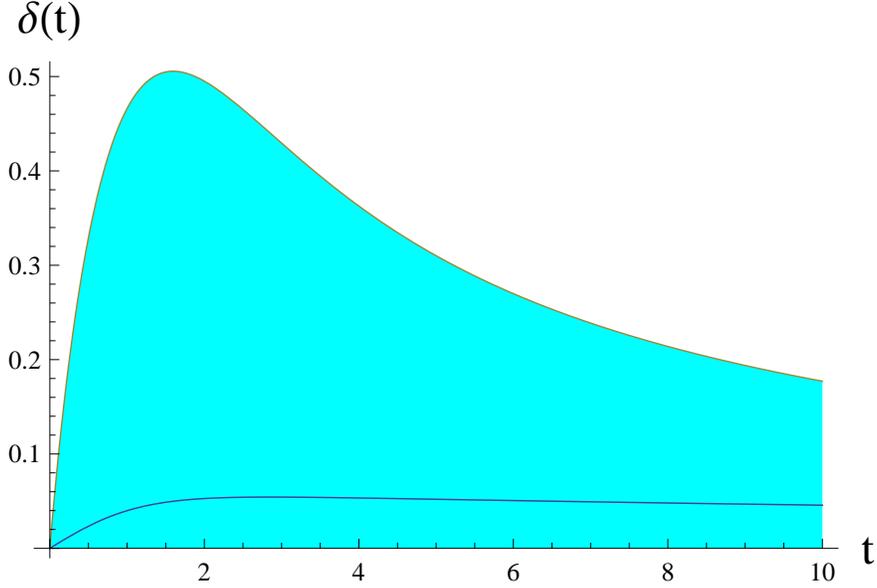}
	\caption{Evolution of dimensionless parameter $\delta$ of cosmological time $t$ for trajectories of type II (for the best fitted values of model parameter together with confidence level at $95\%$). Note that as trajectory in the phase space achieved the state of the pericentrum located in the saddle point, this state is corresponding on the diagram the maximum. Note that the existence of a maximum value of $\delta$ parameter $(\frac{\rho_{m,0}}{\gamma}+t_{max}-t_0)\rho_{m}(t_{max})=2 H(t_{max})$. For the late time $\delta(t)$ function is decreasing function of $t$ and $\rho(\infty)$=0. We choose (s Mpc)/(100$\times$km) as a unit of the cosmological time $t$.}
	\label{fig:fig4}
\end{figure}

Note that the interval of the values of $\rho_{dm}$ is $(0,+\infty)$ or $(0,\rho_{dm}^{\text{max}})$, which depends on a type of trajectories. The evolution of the scale factor $a(t)$ with respect to the cosmological time, for trajectories of type II, is demonstrated in Figure~\ref{fig:fig3}. The function $\delta(t)$, for trajectories of type II, is presented in Figure~\ref{fig:fig4} and the consideration for the maximum of this function is in the form 
\begin{equation}
\left(\frac{\rho_{m,0}}{\gamma}+t_{max}\right)\rho_{m}(t_{\text{max}})=2 H(t_{\text{max}}),
\end{equation}
where $t_{\text{max}}$ is corresponding the value of the cosmological time at the maximum. 

\begin{figure}
	\centering
	\includegraphics[width=0.7\linewidth]{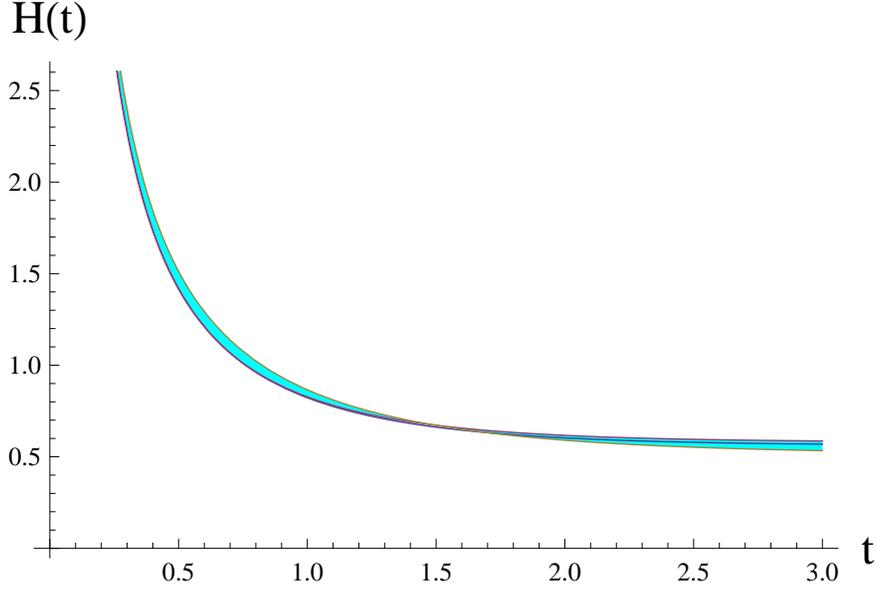}
	\caption{Dependence of Hubble function of trajectories of type II (for the best fitted values of model parameter together with confidence level at $95\%$). For late times $H(t)$ goes to constant values (deS$_+$). The Hubble function $H(t)$ is expressed in [100$\times$km/(s Mpc)]. We choose (s Mpc)/(100$\times$km) as a unit of the cosmological time $t$.}
	\label{fig:fig5}
\end{figure}

\begin{figure}
	\centering
	\includegraphics[width=0.7\linewidth]{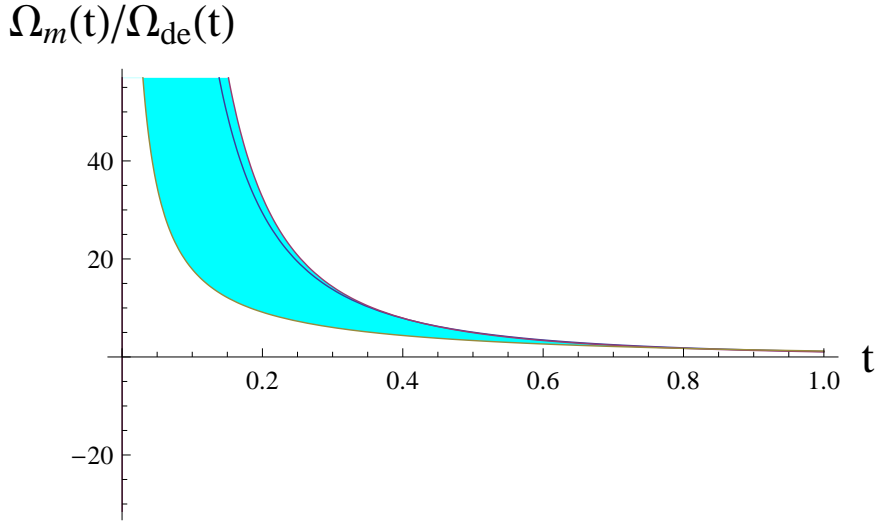}
	\caption{Diagram of relation $\Omega_{m}/\Omega_{de}$ for trajectories of type II (for the best fitted values of model parameter together with confidence level at $95\%$). We choose (s Mpc)/(100$\times$km) as a unit of the cosmological time $t$.}
	\label{fig:fig8}
\end{figure}

\begin{figure}
	\centering
	\includegraphics[width=0.7\linewidth]{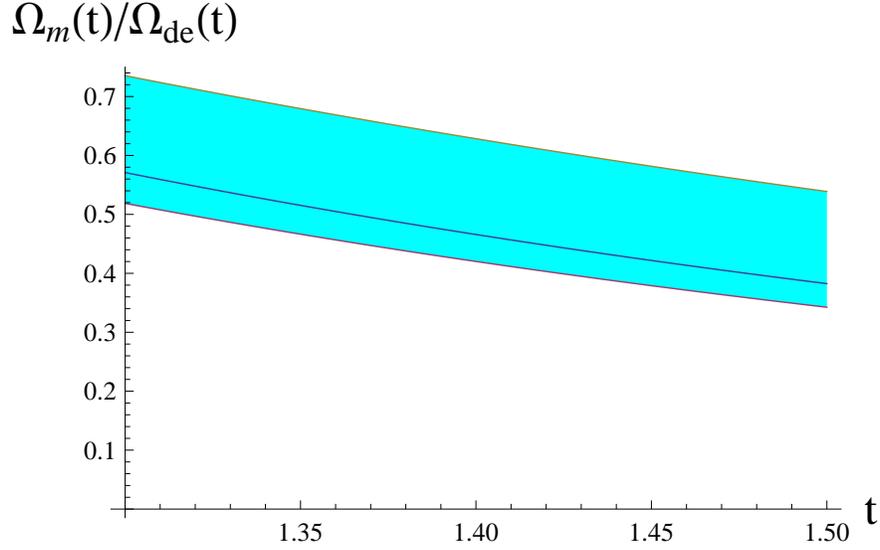}
	\caption{Diagram of relation $\Omega_{m}/\Omega_{de}$ for trajectories of type II for the present epoch. Note that at the present epoch $\rho_{m,0}\propto\rho_{de,0}$ (therefore coincidence problem is solved). We choose (s Mpc)/(100$\times$km) as a unit of the cosmological time $t$. The value of the age of the Universe for the best fit with errors are presented by the dashed lines.}
	\label{fig:fig9}
\end{figure}

\begin{figure}
	\centering
	\includegraphics[width=0.7\linewidth]{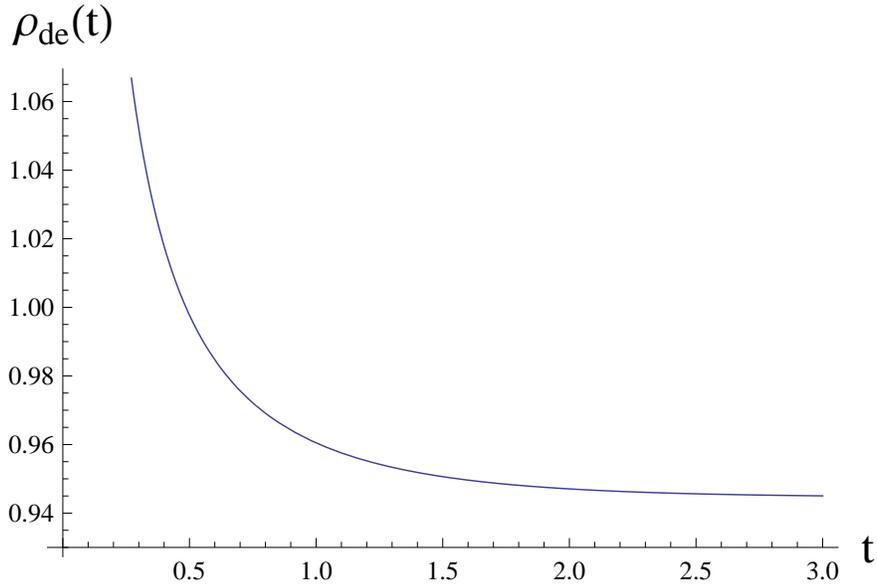}
	\caption{The evolution of dark energy density for the best fitted values of model parameter for trajectories of type II. Dark energy $\rho_{de}$ is expressed in [100$\times$km/(s Mpc)]$^2$. We choose (s Mpc)/(100$\times$km) as a unit of the cosmological time $t$.}
	\label{fig:fig10}
\end{figure}

\begin{figure}
	\centering
	\includegraphics[width=0.7\linewidth]{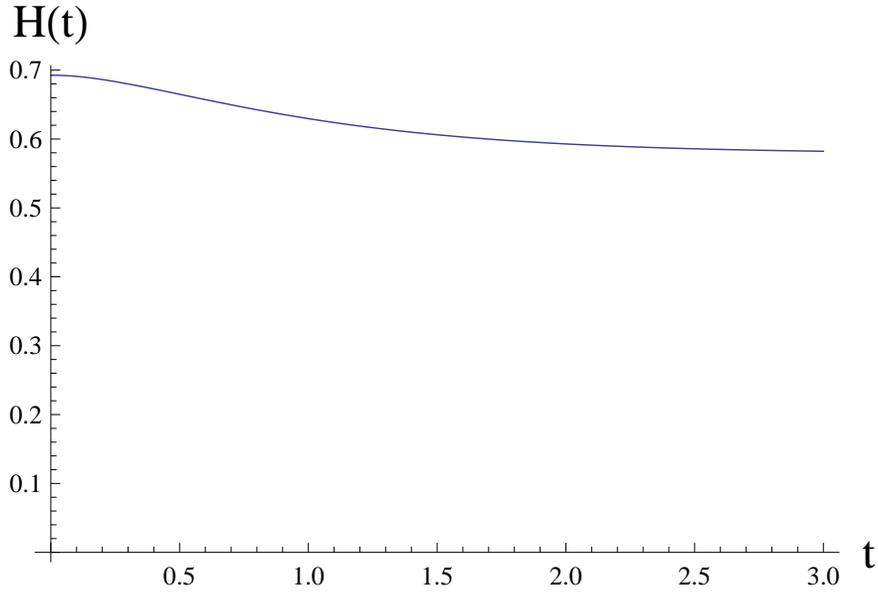}               
	\caption{The relation of $H(t)$ for typical trajectory of type I. The $H(t)$ function is expressed in [100$\times$km/(s Mpc)]. Note that $H(0)$ is finite therefore it is not a singularity. We choose (s Mpc)/(100$\times$km) as a unit of the cosmological time $t$.}
	\label{fig:fig6}
\end{figure}

\begin{figure}
	\centering
	\includegraphics[width=0.7\linewidth]{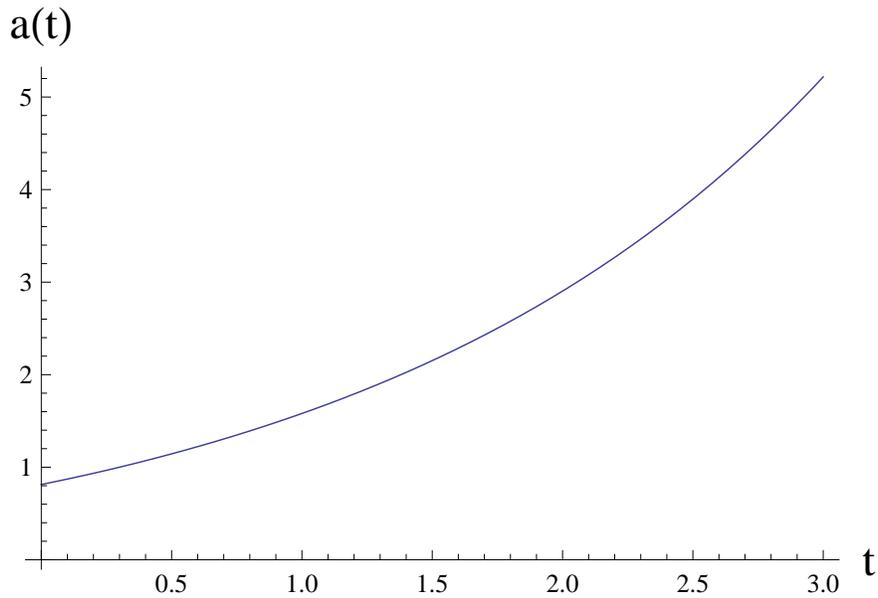}
	\caption{Diagram of $a(t)$ for typical trajectory of type I. Note that $a(0)$ is finite therefore it is not a singularity. We choose (s Mpc)/(100$\times$km) as a unit of the cosmological time $t$.}
	\label{fig:fig7}
\end{figure}

\begin{figure}
	\centering
	\includegraphics[width=0.7\linewidth]{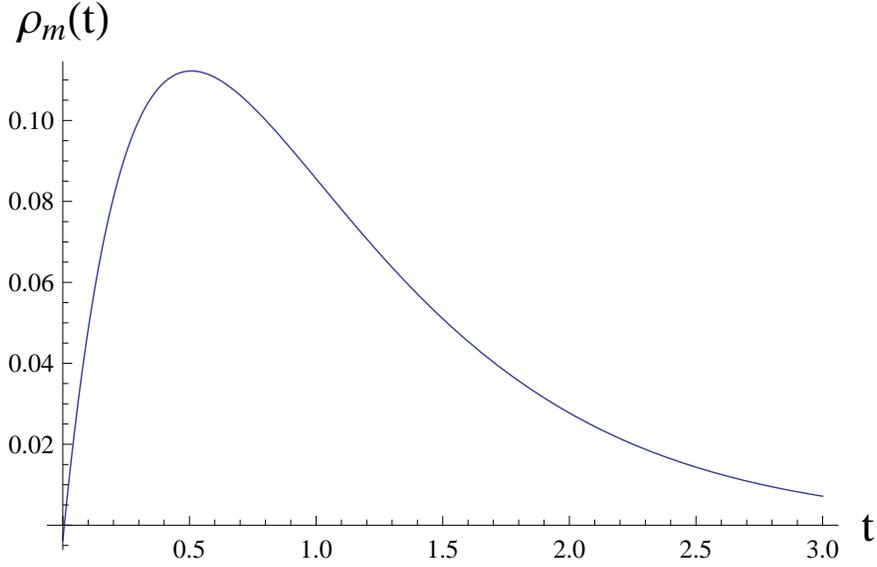}
	\caption{Diagram of $\rho_m (t)$ for typical trajectory of type I. Note that $\rho_{m}(0)$ is equal zero therefore it is not a singularity. Matter $\rho_{m}$ is expressed in [100$\times$km/(s Mpc)]$^2$. We choose (s Mpc)/(100$\times$km) as a unit of the cosmological time $t$.}
	\label{fig:fig12}
\end{figure}

\begin{figure}
	\centering
	\includegraphics[width=0.7\linewidth]{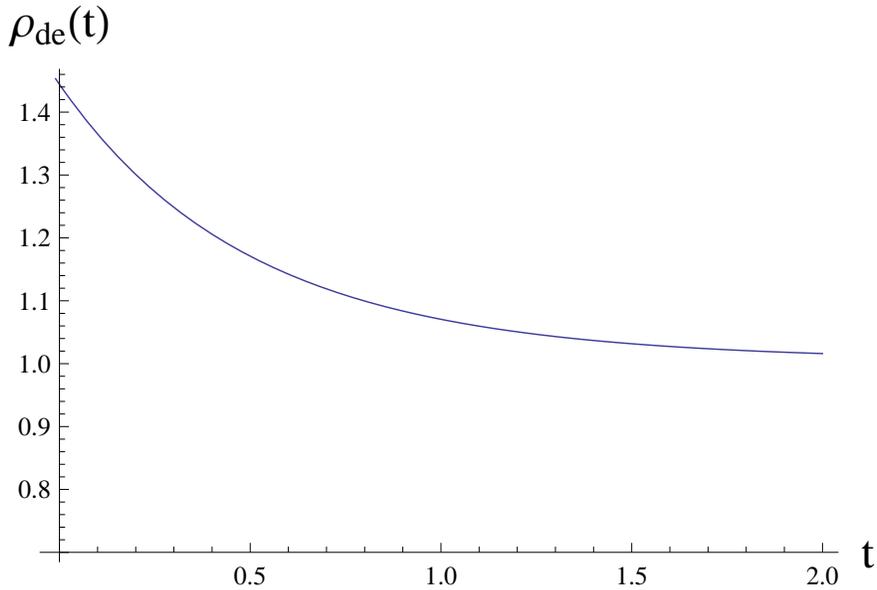}
	\caption{Diagram of $\rho_{de} (t)$ for typical trajectory of type I. Note that $\rho_{de}(0)$ is finite therefore it is not a singularity. Dark energy $\rho_{de}$ is expressed in [100$\times$km/(s Mpc)]$^2$. We choose (s Mpc)/(100$\times$km) as a unit of the cosmological time $t$.}
	\label{fig:fig13}
\end{figure}

\begin{figure}
	\centering
	\includegraphics[width=0.7\linewidth]{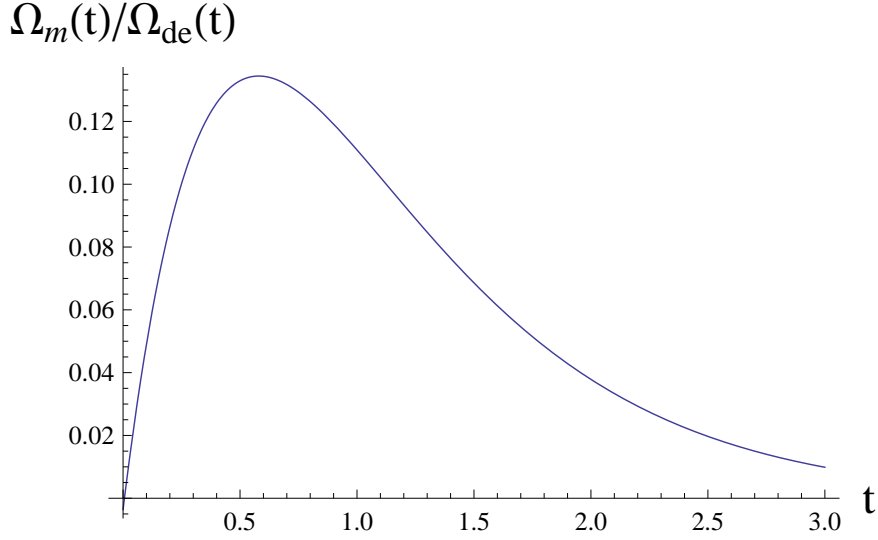}
	\caption{Diagram of $\Omega_m (t)/\Omega_{de} (t)$ for typical trajectory of type I. We choose (s Mpc)/(100$\times$km) as a unit of the cosmological time $t$.}
	\label{fig:fig14}
\end{figure}

\begin{figure}
	\centering
	\includegraphics[width=0.7\linewidth]{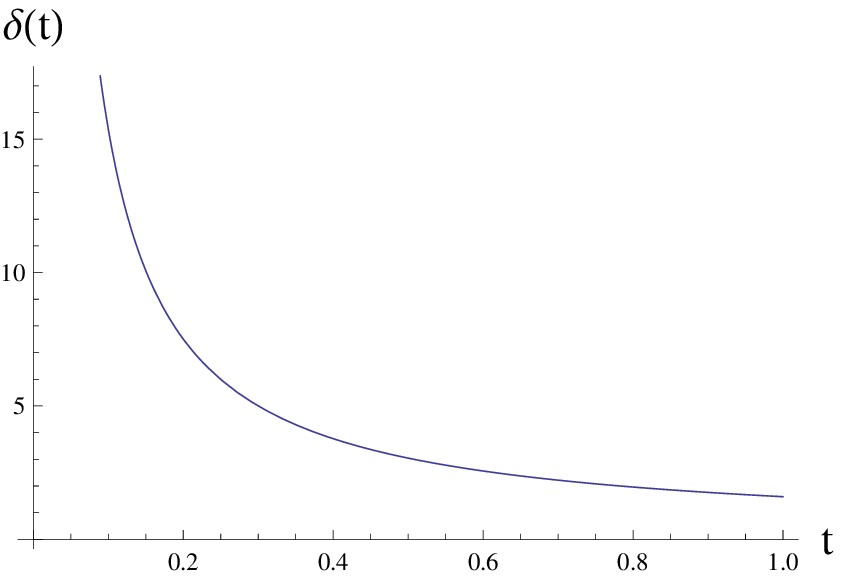}
	\caption{Diagram of $\delta (t)$ for typical trajectory of type I. We choose (s Mpc)/(100$\times$km) as a unit of the cosmological time $t$.}
	\label{fig:fig15}
\end{figure}

The Hubble function $H(t)$, for trajectories of type II, is presented in Figure~\ref{fig:fig5}. Note that the Hubble function in the late time is constant. The evolution of $\rho_{de}$, for trajectories of type II, is shown in Figure~\ref{fig:fig10} and for late time $\rho_{de}$ is going toward constant value. The evolution of $\Omega_{m}/\Omega_{de}$, for trajectories of type II, is demonstrated in Figures \ref{fig:fig8} and \ref{fig:fig9}. For the trajectories of type I, the functions $H(t)$, $a(t)$, $\rho_{dm}$, $\rho_{de}$, $\Omega_{m}/\Omega_{de}$ and $\delta(t)$ are presented in Figures \ref{fig:fig6}, \ref{fig:fig7}, \ref{fig:fig12}, \ref{fig:fig13}, \ref{fig:fig14}, \ref{fig:fig15}. These figures show that there are two distinct behavior of trajectories of type I and II. While the trajectory of type II represents matter dominating model with a singularity the trajectories of type I represents the model without an initial singularity.

\begin{table}
	\caption{Critical points for dynamical system (\ref{poincare1})-(\ref{poincare2}), their type and cosmological interpretation.}
	\label{table:1}
	\begin{center}
		\begin{tabular}{lllllll} \hline
			\scriptsize{No.} & \scriptsize{critical point} &\scriptsize{type of critical}&\scriptsize{type of universe} &\scriptsize{dominating part in} &\scriptsize{$H(t)$} & \scriptsize{$a(t)$}\\& & \scriptsize{point} & &\scriptsize{the Friedmann equation}& &\\ \hline \hline
			1 & \scriptsize{$X_0=0$, $\Delta_0=0$} &\scriptsize{saddle} & \scriptsize{de Sitter universe} &\scriptsize{cosmological}&\scriptsize{$H(t)=\sqrt{\frac{\Lambda_{bare}}{3}}$}&\scriptsize{$a(t)\propto e^{\sqrt{\frac{\Lambda_{bare}}{3}}t}$}\\
			& & &\scriptsize{without diffusion effect} &\scriptsize{constant}& &\\
			2 & \scriptsize{$X_0=\sqrt{2/11}$, $\Delta_0=3/\sqrt{22}$ }&\scriptsize{saddle} & \scriptsize{scaling universe} &\scriptsize{matter and}&\scriptsize{$H(t)=(t-t_0)^{-1}$}& \scriptsize{$a(t)\propto(t-t_0)$}\\ &&&\scriptsize{($\rho_{m}\propto\rho_{de}$)}&\scriptsize{dark energy}&&\\
			3 & \scriptsize{$X_0=1/\sqrt{2}$, $\Delta_0=0$ }&\scriptsize{unstable node} & \scriptsize{Einstein-de Sitter}&\scriptsize{matter}&\scriptsize{$H(t)=\frac{2}{3}(t-t_0)^{-1}$}&\scriptsize{$a(t)\propto(t-t_0)^{2/3}$}\\ & & & \scriptsize{universe} & & &\\
			4 & \scriptsize{$X_0=1$, $\Delta_0=0$ }&\scriptsize{stable node} & \scriptsize{static} &\scriptsize{matter and}&\scriptsize{$H(t)=0$}&\scriptsize{$a(t)=const$}\\ &&&\scriptsize{universe}&\scriptsize{running dark energy}&&\\
			5 & \scriptsize{$X_0=4/5$, $\Delta_0=-3/5$ }&\scriptsize{saddle} & \scriptsize{static} &\scriptsize{matter and}&\scriptsize{$H(t)=0$}&\scriptsize{$a(t)=const$}\\ &&&\scriptsize{universe}&\scriptsize{running dark energy}&&\\
			6 & \scriptsize{$X_0=0$, $\Delta_0=1$ }&\scriptsize{unstable node} & \scriptsize{de Sitter universe}&\scriptsize{running dark energy}&\scriptsize{$H(t)=\sqrt{\frac{\rho_{de}(0)}{3}}$}&\scriptsize{$a(t)\propto e^{\sqrt{\frac{\rho_{de}(0)}{3}}t}$}\\ & & & \scriptsize{with diffusion effect} & & &\\
			7 & \scriptsize{$X_0=0$, $\Delta_0=-1$} &\scriptsize{stable node} & \scriptsize{de Sitter universe}&\scriptsize{running dark energy}&\scriptsize{$H(t)=-\sqrt{\frac{\Lambda_{bare}}{3}}$}&\scriptsize{$a(t)\propto e^{-\sqrt{\frac{\Lambda_{bare}}{3}}t}$}\\ & & & \scriptsize{with diffusion effect} & & &\\
			\\ \hline
		\end{tabular}
	\end{center}
\end{table}

\section{Generalized diffusion cosmology}

Dynamical system methods are especially suitable in investigation of dynamics of both fluids, dark energy and dark matter. Presented here dynamical system approach to the study DM-DE interaction in diffusion cosmology can be simply generalized to the case when both dark energy and dark matter satisfy a general form of the equation of state
\begin{align}
p_{de}&=w \rho_{de}, \\
p_{dm}&=\tilde{w} \rho_{dm}, \\
p_b&=0,
\end{align}
where $w$ and $\tilde{w}$ are constant coefficients equation of state for dark energy and matter respectively. Then the continuity equations for baryonic and dark matter and dark energy are presented by
\begin{align}
	\dot\rho_{dm}&=-3(1+\tilde{w})H\rho_{dm}+\gamma a^{-3},\label{darkmatter3} \\
	\dot\rho_{de}&=-3(1+w)H\rho_{de}-\gamma a^{-3},\label{darkenergy3} \\
	\dot\rho_{b}&=-3H\rho_{b}.\label{baryon} \\
\end{align}

The corresponding dynamical system assumes the form of a 3-dimensional autonomous dynamical system
\begin{align}
\frac{dx}{d\ln a}&=3x\left[(1+\tilde{w})(x-1)+(1+w)y+\frac{z}{3}\right],\\
\frac{dy}{d\ln a}&=3y[(1+w)(y-1)+(1+\tilde{w})x]-xz,\\
\frac{dz}{d\ln a}&=z\left[3\tilde{w}-z+\frac{3}{2}[(1+\tilde{w})x+(1+w)y]\right],
\end{align}
where we choose state variables $x =\Omega_m$, and $y=\Omega_{de}$ and $z =\delta$ like in previous considered case.

Because $1=x+y$ the above dynamical system reduces to
\begin{align}
\frac{dx}{d\ln a}&=3x\left[(\tilde{w}-w)(x-1)+\frac{z}{3}\right],\label{dyn4}\\
\frac{dz}{d\ln a}&=z\left[3\tilde{w}-z+\frac{3}{2}[(1+\tilde{w})x+(1+w)(1-x)]\right].\label{dyn5}
\end{align}
Critical points of dynamical system (\ref{dyn4})-(\ref{dyn5}) are completed in Table \ref{table:2}. Especially there is an interesting critical point inside the admissible region $D=\{(x,z)\colon x\geq 0, z\geq 0\}$ representing scaling solution: $\rho_{dm}\propto\rho_{de}$. It is a saddle fixed point in the phase space $D$. This critical point is important in the context of solution of cosmic coincidence problem as well as scaling solution in the context of quintessence idea.

\begin{table}
	\caption{Critical points for dynamical system (\ref{dyn4})-(\ref{dyn5}), their positions, types and cosmological interpretation.}
	\label{table:2}
	\begin{center}
		\begin{tabular}{lll} \hline
			No. & Critical point & type of the universe \\ \hline \hline
			1& $x_0=0$, $z_0=0$ & de Sitter universe \\
			&& without diffusion\\
			2& $x_0=1$, $z_0=0$ & Einstein-de Sitter \\
			3& $x_0=-\frac{1+3w}{3(\tilde{w}-w)}$, $z_0=1 + 3 \tilde{w}$ & scaling universe $(\rho_m\propto\rho_{de})$ \\
			4& $x_0=0$, $z_0=3/2 (1 + 2 \tilde{w} + w)$ & de Sitter universe \\ && with diffusion\\
			\\ \hline
		\end{tabular}
	\end{center}
\end{table}

Above system possesses critical points on the planes of coordinate system or inside the phase space $D=\{(x,z)\colon x,z\geq 0\}$. Of course system under consideration is restricted to the submanifold $x+y=1$, because the constrain condition $\Omega_m+\Omega_{de}=1$.

The behavior of trajectories of dynamical system (\ref{dyn4})-(\ref{dyn5}) depends on the values of parameters $w$, $\tilde{w}$. By choosing different values of these parameters one can study how phase space structure changes under change of values of parameters. The equivalence of the phase portraits is establish following homeomorphism preserving direction of time along the trajectories. If there exist value of parameter for which phase is not topologically equivalent, then such value is bifurcation value.

The stability of critical points depends on the linearization matrix. At the critical point (3), the linearization matrix has the following form
\begin{equation}
A=\left(\begin{array}{cc}
\frac{\partial f_x(x,z)}{\partial x}|_{x_0,z_0} & \frac{\partial f_x(x,z)}{\partial z}|_{x_0,z_0}\\ \frac{\partial f_z(x,z)}{\partial x}|_{x_0,z_0} & \frac{\partial f_z(x,z)}{\partial z}|_{x_0,z_0}
\end{array}\right)=\left(\begin{array}{cc}
-1-3w & \frac{1+3w}{-3\tilde{w}+3w}\\ \frac{3}{2}(1+3\tilde{w})(\tilde{w}-w) & -1-3\tilde{w}
\end{array}\right),\label{matrix}
\end{equation}
where $f_x(x,z)$ and $f_z(x,z)$ are the right sides of equations (\ref{dyn4}) and (\ref{dyn5}) and $x_0$ and $z_0$ are coordinates of critical point (3) (see table \ref{table:2}).

The determinant of matrix (\ref{matrix}) can be expressed by the formula
\begin{equation}
\det A=\frac{3}{2}(1+3\tilde{w})(1+3w) \label{det}
\end{equation}
and the trace of matrix (\ref{matrix}) is described by
\begin{equation}
\text{tr } A=-2-3(\tilde{w}+w).\label{trace}
\end{equation}
Therefore the critical point (3) is stable when $w+\tilde{w}>-2/3$.

\begin{figure}
	\centering
	\includegraphics[width=0.7\linewidth]{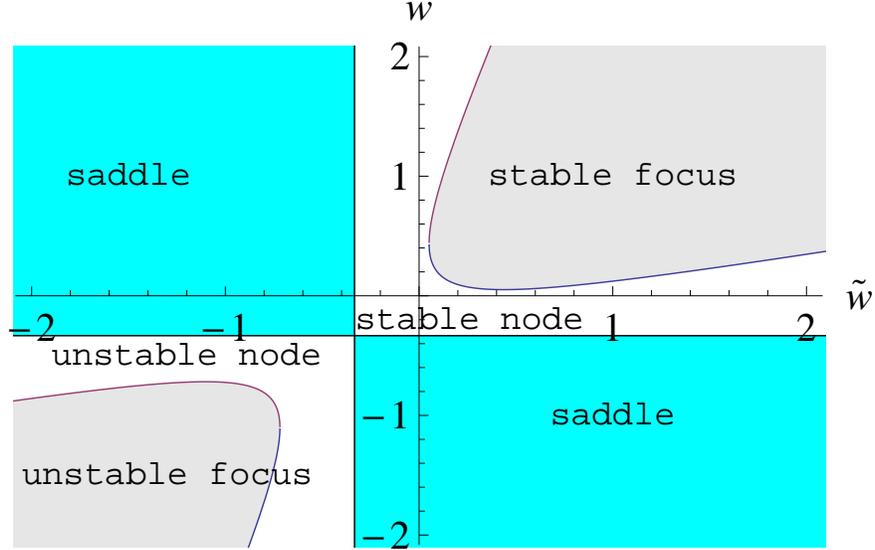}
	\caption{Diagram of stability of critical point (3), depending on $w$ and $\tilde{w}$. In the gray domains there is the focus type of critical point and the boundaries of this domains is given by the lines $w=\frac{1}{3}(1+6\tilde{w}+\sqrt{-1+9\tilde{w}(2+3\tilde{w})})$ and $w=\frac{1}{3}(1+6\tilde{w}-\sqrt{-1+9\tilde{w}(2+3\tilde{w})})$. In the blue regions there are the saddle type of critical point and is limited by lines $w=-1/3$ and $\tilde{w}=-1/3$. In the white top and bottom regions there are the stable and unstable nodes, respectively.}
	\label{fig:fig16}
\end{figure}

The characteristic equation for matrix $A$ at critical point (3) is in the following form
\begin{equation}
\lambda^2-\text{tr }A\lambda+\det{A}=\lambda^2+(2+3\tilde{w}+3w)\lambda+\frac{3}{2}(1+3\tilde{w})(1+3w)=0.\label{chara}
\end{equation}
From the characteristic equation (\ref{chara}), we can obtain the eigenvalues for critical point (3) (see Table \ref{table:2}). In Figure \ref{fig:fig16} we demonstrate the stability of critical point (3), depending on $w$ and $\tilde{w}$.

The linearized equation (\ref{dyn4})-(\ref{dyn5}) at critical point (3) is given by the following formulas
\begin{equation}
\begin{array}{c}
(x-x_0)'=A_{11}(x-x_0)+A_{12}(z-z_0)\\
=(-1-3w)\left(x+\frac{1+3w}{3(\tilde{w}-w)}\right)+\left(\frac{1+3w}{-3\tilde{w}+3w}\right)(z-1 - 3 \tilde{w}),
\end{array}
\end{equation}
\begin{equation}
\begin{array}{c}
(z-z_0)'=A_{21}(x-x_0)+A_{22}(z-z_0)\\
=\frac{3}{2}(1+3\tilde{w})(\tilde{w}-w)\left(x+\frac{1+3w}{3(\tilde{w}-w)}\right)+\left(-1-3\tilde{w}\right)(z-1 - 3 \tilde{w}),
\end{array}
\end{equation}
where $x_0=-\frac{1+3w}{3(\tilde{w}-w)}$ and $z_0=1 + 3 \tilde{w}$.
The solutions of the above equations are presented by formulas
\begin{equation}
x=C_1 a^{(-2-3\tilde{w}-3w-\alpha)/2}(a^{\alpha}+C_2)-\frac{1+3w}{3(\tilde{w}-w)},
\end{equation}
\begin{equation}
z=C_1\frac{3(\tilde{w}-w)}{2+6w}a^{(-2-3\tilde{w}-3w-\alpha)/2}((3\tilde{w}-3w-\alpha)a^{\alpha}+C_2 (3\tilde{w}-3w+\alpha))+1 + 3 \tilde{w},
\end{equation}
where $\alpha=\sqrt{-1+9\tilde{w}^2+9w^2-(1+6\tilde{w})(1+6w)}$.

It is interesting to check how structure of phase space changes under changing coefficient equation of state for dark matter from 0 (cold dark matter) to $\tilde{w}=1/3$ (hot dark matter).

Results of dynamical investigation show that structure of the phase space is preserved under changes of the model parameter. Let us considered some details.

\section{Diffusion cosmology with the hot relativistic dark matter}
		
In this section we consider the case with relativistic dark matter ($\tilde{w}=1/3$) and $w=-1$. Then the equation of state for dark matter is in form $p_{dm}=\frac{1}{3}\rho_{dm}$, where $p_{dm}$ is the pressure of dark matter. We get the following equations
	\begin{equation}
	x'=x(-4+z+4x),\label{dyn6}
	\end{equation}
	\begin{equation}
	z'=z(1-z+2x).\label{dyn7}
	\end{equation}
	
\begin{figure}
		\centering
		\includegraphics[width=0.7\linewidth]{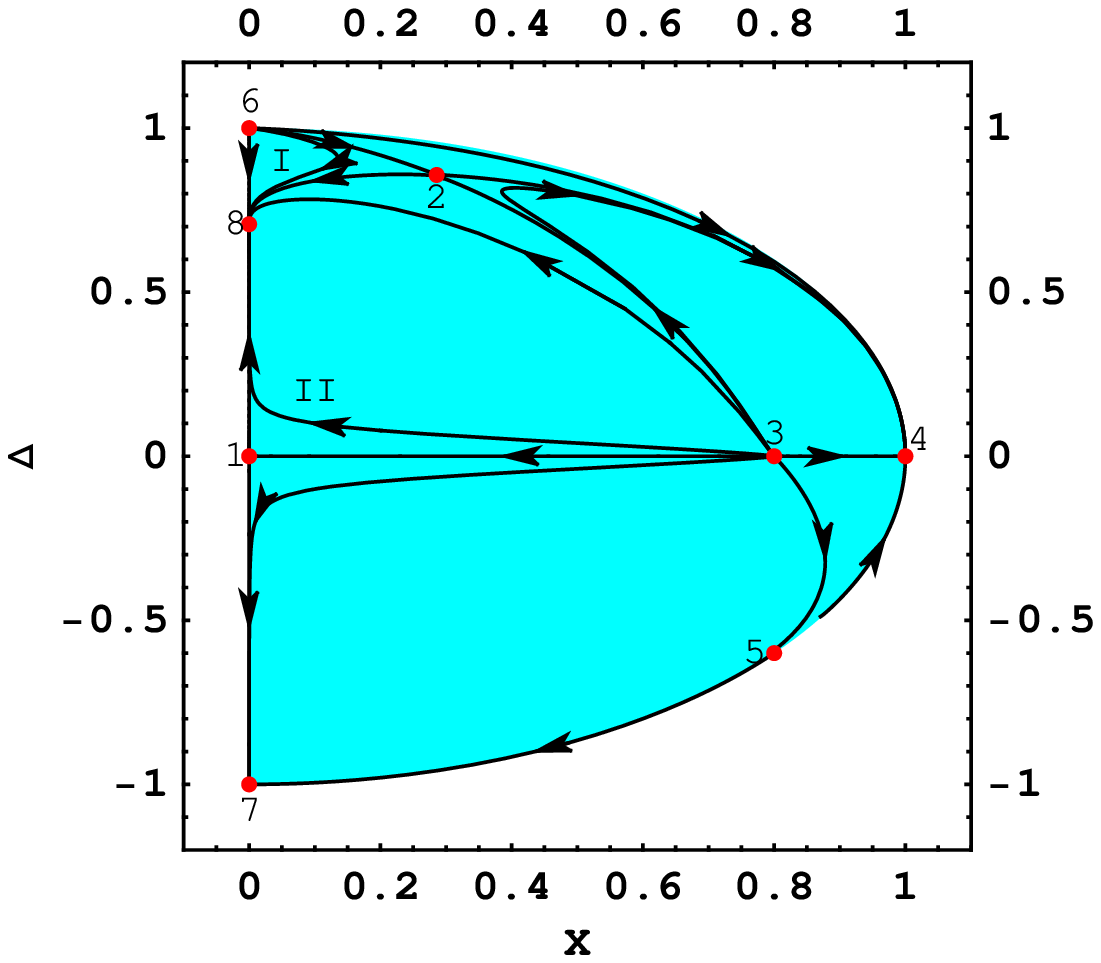}
		\caption{Phase portrait of dynamical system (\ref{dyn4})-(\ref{dyn5}). Note that trajectories for the $\Delta<0$ represent solutions with the negative value of $H$. From the cosmological point of view trajectories representing expanding models with $\Delta>0$ are physical. Critical points (1) represents the de Sitter universe without diffusion effect. Critical point (2) represents the scaling universe. Critical point (3) represents the Einstein-de Sitter universe. Critical points (4) and (5) represent the static universe. Critical point (6) and (8) represent the de Sitter universe with diffusion effect.}
		\label{fig:fig1}
	\end{figure}

	We can analyze the critical points in the infinity. In this case we use the Poincar{\'e} sphere. Let $X=\frac{x}{\sqrt{x^2+\delta^2}}$, $\Delta=\frac{\delta}{\sqrt{x^2+\delta^2}}$. For variables $X$ and $\Delta$, we get dynamical system
	\begin{equation}
	X'=X\left[-\Delta^2(\sqrt{1-X^2-\Delta^2}+\frac{3}{2}X-\Delta)+(1-X^2)
	(3X+\Delta-4\sqrt{1-X^2-\Delta^2})\right],\label{poincare3}
	\end{equation}
	\begin{equation}
	\Delta'=\Delta\left[(1-\Delta^2)(\sqrt{1-X^2-\Delta^2}+\frac{3}{2}X-\Delta)-X^2
	(3X+\Delta-4\sqrt{1-X^2-\Delta^2})\right],\label{poincare4}
	\end{equation}
	where $'\equiv \sqrt{1-X^2-\Delta^2}\frac{d}{d\tau}$. Critical points, for the above equation, are presented in table \ref{table:3}. The phase portrait for the dynamical system (\ref{poincare3})-(\ref{poincare4}) are demonstrated in Figure \ref{fig:fig1}.
	
	\begin{table}
		\caption{Critical points for dynamical system (\ref{poincare3})-(\ref{poincare4}), their type and cosmological interpretation.}
		\label{table:3}
		\begin{center}
			\begin{tabular}{llll} \hline
				No. & critical point & type of critical point & type of universe \\ \hline \hline
				1 & $X_0=0$, $\Delta_0=0$ & saddle & de Sitter universe without diffusion effect \\
				2 & $X_0=2/7$, $\Delta_0=6/7$ & saddle & scaling universe \\
				3 & $X_0=4/5$, $\Delta_0=0$ & unstable node & Einstein-de Sitter universe\\
				4 & $X_0=1$, $\Delta_0=0$ & stable node & static universe\\
				5 & $X_0=4/5$, $\Delta_0=-3/5$ & saddle & static universe\\
				6 & $X_0=0$, $\Delta_0=1$ & unstable node & de Sitter universe with diffusion effect \\
				7 & $X_0=0$, $\Delta_0=-1$ & stable node & de Sitter universe with diffusion effect \\
				8 & $X_0=0$, $\Delta_0=1/\sqrt{2}$ & stable node & de Sitter universe with diffusion effect \\
				\\ \hline
			\end{tabular}
		\end{center}
	\end{table}
	
It is interesting to see how different solutions with and without an initial singularity are distributed in the phase space. To answer this question it would be useful to consider a phase space structure of the models under consideration. For this aim we reduce the dynamics to the form of an autonomous 2D dynamical system. In such a system the state variables are dimensionless parameters: the density parameter for radiation dark matter and the parameter $\delta$ characterizing the rate of energy transfer to the dark matter sector.
	
The main advantage of visualization global dynamics on the phase portrait is the possibility to see all solution of the system admitted for all initial conditions. On the phase portrait there is the geometric representation of evolutional paths of both solution types. Critical points are representing asymptotic states of the system, i.e. stationary states. In order trajectories joining different critical points are representing the evolution of the system.
	
Similarly to dynamical investigations presented in our previous paper \cite{Haba:2016swv} we added to the plane circle at infinity via the construction of the Poincar{\'e} sphere. Hence we obtain a compact phase space and consequently a global phase portrait. In Fig. \ref{fig:fig1} we have identified linear solutions without the initial singularity as the representing by saddle critical point (2). In the phase portrait there are critical points at a finite domain as well as located on the boundary at infinity.
	
Note that the phase portrait has no symmetry with respect to the x-axis. Critical point (6) is representing an expanding stationary de Sitter type solution determined by diffusion effects. We denote as typical trajectories starting from this critical point and going toward the de Sitter empty universe. This type of trajectories we called trajectories of type I. In the phase portrait there are also present trajectories of type II. These trajectories are starting from the Einstein-de Sitter universe with the initial singularity and coming toward the de Sitter universe labeled as critical point (8). 
	
Looking at the phase portrait one can observe that only critical points of type unstable and stable node (global attractors and global repellers) and saddle appear in the phase space. Therefore the model obtained is structurally stable, i.e. any small change of its r.h.s does not disturb the global phase portrait. Physically this means that corresponding model is realistic. Mathematically this fact has a nice interpretation in the context of the Peixoto theorem \cite{Perko:2001de} that they are generic in this sense because they form open and dense subsets in the space of dynamical systems on the plane.
	
\section{Conclusion}

The standard cosmological model ($\Lambda$CDM model) is widely accepted but it has still some problems, namely the cosmological constant problem and the coincidence problem.
In the standard cosmological model ($\Lambda$CDM model) it is assumed that all fluids are non-interacting. This implies that the densities of baryonic matter and dark matter are scaling with respect to redshift as $\rho_{\text{dm},0} a(t)^{-3}$ and dark energy has constant density. In this paper we construct a cosmological model in which it is assumed that the process of interaction between sectors of dark matter and dark energy is continuous. Relativistic diffusion describes the transfer of energy to the sector of dark matter. This effect is described by the running cosmological constant and the modification of the standard scaling law of the dark matter density to the form $\rho_{\text{dm},0} a(t)^{-3} + \gamma t a(t)^{-3}$. The dynamics of this model is studied for possible explanations of cosmological puzzles: the cosmological conatant problem and the coincidence problem.

In the context of the coincidence problem, our model can explain the present ratio of $\rho_{m}$ to $\rho_{de}$, which is equal $0.4576^{+0.1109}_{-0.0831}$ at a 2$\sigma$ confidence level. In our model, the canonical scaling law of dark matter $(\rho_{dm,0}a^{-3}(t))$ is modified by an additive $\epsilon(t)=\gamma t a^{-3}(t)$ to the form $\rho_{dm}=\rho_{dm,0}a^{-3}(t)+\epsilon(t)$.

The analysis of the time dependence of density of dark energy and dark matter, we conclude that the value of effective energy of vacuum runs from an infinite value to a constant value, and the delta amendment to the scaling law goes from zero to zero and being different from zero in a long intermediate period. This characteristic type of behavior is controlled by the diffusion effect.

The paper presents a detailed study of the behavior of a state of the system represented by the state variables $(x, \delta)$. In this context, it was natural to consider the diffusion mechanism which controls the change of the ratio of both energy densities and the very dynamics of this process remains in analogy to the description of population changes of competing species \cite{Perez:2013zya}. A crucial role plays the saddle critical point $H a=const$, which is a scaling type of the solution ($\rho_{m}\propto\rho_{de}$). The position of this point cannot be disturbed by a small perturbation (structurally stable point as well as whole system).

\acknowledgments{The work was supported by the grant NCN DEC-2013/09/B/ST2/03455. The authors thank prof. Z. Haba and A. Krawiec for remarks and comments.}


%

\end{document}